\documentclass[12pt]{article}
\usepackage{latexsym}
\usepackage{amssymb,amsfonts,amsmath}
\usepackage{graphicx} 
\usepackage{indentfirst}
\usepackage{bbm}
\usepackage{amssymb}
\usepackage{verbatim}
\usepackage{amsmath, amsthm,amssymb}
\usepackage{mathrsfs}
\usepackage{hyperref}
\usepackage{amsfonts}
\usepackage{dsfont}
\newcommand {\cD}{{\cal D}}
\newcommand {\cE}{{\cal E}}

\newcommand {\cG}{{\cal G}}
\newcommand {\cH}{{\cal H}}

\newcommand {\cL}{{\cal L}}

\newcommand {\cN}{{\cal N}}

\newcommand {\cR}{{\cal R}}

\newcommand {\cT}{{\cal T}}

\newcommand {\cW}{{\cal W}}

\def\a{\alpha}
\def\b{\beta}

\def\d{\delta}

\def\f{\phi}
\def\g{\gamma}
\def\G{\Gamma}

\def\j{\psi}

\def\l{\lambda}

\def\o{\omega}

\def\q{\theta}
\def\r{\rho}
\def\s{\sigma}
\def\t{\tau}

\def\x{\xi}
\def\z{\zeta}

\def\J{\Psi}
\def\L{\Lambda}
\def\O{\Omega}

\def\S{\Sigma}
\def\U{\Upsilon}

\def\ri{{\rm i}}
\def\re{{\rm e}}

\newcommand{\ad}{{\dot{\alpha}}}                           %new
\newcommand{\bd}{{\dot{\beta}}}                            %new
\newcommand{\ve}{\varepsilon}                            %new
\newcommand{\cDB}{{\bar\cD}}                            %new

\newcommand{\pa}{\partial}                           %new
\newcommand{\hf}{\frac12}
\newcommand{\be}{\begin{equation}}
\newcommand{\ee}{\end{equation}}
\newcommand{\bea}{\begin{eqnarray}}
\newcommand{\eea}{\end{eqnarray}}
\newcommand{\non}{\nonumber}
\newcommand{\ba}{\begin{array}}
\newcommand{\ea}{\end{array}}

\newcommand{\1}{{\underline{1}}}
\newcommand{\2}{{\underline{2}}}

\newcommand{\bm}[1]{\mbox{\boldmath$#1$}}

\def\double #1{#1{\hbox{\kern-2pt $#1$}}}

\newcommand{\gd}{{\dot\g}}

\newcommand{\sSL}{\mathsf{SL}}

\newcommand{\sU}{\mathsf{U}}

\newcommand{\bsubeq}{\begin{subequations}}
\newcommand{\esubeq}{\end{subequations}}

\newcommand{\rd}{\mathrm d}
\begin{document}
%%%%%%%%%%%%%%%%
%%%%%%%%%%%%%%%%

\begin{center}
{\Large \bf 
Superconformal vector multiplet self-couplings \\
and generalised  Fayet-Iliopoulos terms}
\end{center}

\begin{center}
{\bf Sergei M. Kuzenko} \\
\vspace{5mm}

\footnotesize{
{\it Department of Physics M013, The University of Western Australia\\
35 Stirling Highway, Crawley W.A. 6009, Australia}}  
~\\
\vspace{2mm}
\end{center}

\begin{abstract}
\baselineskip=14pt
As an extension of the recent construction of generalised Fayet-Iliopoulos terms
in ${\cal N}=1$ supergravity given in \cite{K18}, we present self-interactions for a vector multiplet coupled to conformal supergravity. 
They are used to construct new models for spontaneously broken local supersymmetry.
\end{abstract}

\renewcommand{\thefootnote}{\arabic{footnote}}

%%%%%%%%%%%%%%%%%%%%%%%%%%%%%%%%%%%%%%%%%%%%%%%%%%%%%%
%%%%%%%%%%%%%%%%%%%%%%%%%%%%%%%%%%%%%%%%%%%%%%%%%%%%%%

Recently, 
a one-parameter family of generalised  Fayet-Iliopoulos (FI) terms in supergravity
were proposed \cite{K18}, including the one discovered in \cite{CFTVP},
with the crucial property that no gauged 
$R$-symmetry is required, unlike the standard  FI term 
\cite{FI} lifted to supergravity \cite{Freedman,BFNS}.
These generalised FI terms make use of composite super-Weyl
primary multiplets ${\mathbb V}_{(n)} $, with $n$ a real parameter, 
which are  constructed from an Abelian vector multiplet coupled 
to $\cN=1$ conformal supergravity.\footnote{We make use of the superspace formulation for conformal supergravity described in the Appendix.} 
As usual, the vector multiplet
is described using a real scalar prepotential $V$  defined modulo gauge transformations
\bea
\d_\l V = \l +\bar \l~, \qquad \bar \cD_\ad \l =0~.
\label{gaugefreedom}
\eea
The prepotential is chosen to be super-Weyl inert, $\d_\s V=0$. 
It was assumed in \cite{K18}  that the top component ($D$-field) of $V$ 
is nowhere vanishing. 
In terms of  the gauge-invariant covariantly chiral field strength \cite{FZ,WZ}
\bea
W_\a := -\frac{1}{4} (\bar \cD^2 - 4R) \cD_\a V~,\qquad \bar \cD_\bd W_\a=0~,
\label{W}
\eea
this assumption means that the real scalar
$\cD W:=\cD^\a W_\a =\bar \cD_\ad \bar W^\ad$ is nowhere vanishing.\footnote{We 
follow the notation and conventions of \cite{Ideas}.}
The lowest component of $\cD W$ is proportional to the top component
of $V$, see eq.\,\eqref{19}.
In the special case $n=0$, the composite ${\mathbb V} \equiv {\mathbb V}_{(0)} $
is derived as a by-product of the $\cN=1$ Goldstino superfield 
construction proposed in  \cite{KMcAT-M}, and it reads  
\bea
{\mathbb V} := - 4 \frac{W^2 \bar W^2}{(\cD W)^3}~, \qquad 
W^2 := W^\a W_\a~.
\eea
In general,  ${\mathbb V}_{(n)} $ is defined to have the form \cite{K18}
\bea
{\mathbb V}_{(n)} := \Big[ 
\frac{\cD^2 W^2 \bar \cD^2 \bar W^2 }{(\cD W)^{4}}
\Big]^n \,{\mathbb V} ~.
\eea
It has the following   properties: 
\begin{enumerate} 
\item ${\mathbb V}_{(n)} $ satisfies the nilpotency conditions
\bea
{\mathbb V}_{(n)}{\mathbb V}_{(n)}=0~, 
\qquad
{\mathbb V}_{(n)} \cD_A \cD_B {\mathbb V}_{(n)} =0~,
\qquad
{\mathbb V}_{(n)} \cD_A \cD_B \cD_C {\mathbb V}_{(n)}=0~.
\label{5}
\eea
\item ${\mathbb V}_{(n)} $ is gauge invariant, $\d_\l {\mathbb V}_{(n)} =0$. 
\item
${\mathbb V}_{(n)} $ is super-Weyl inert, $\d_\s {\mathbb V}_{(n)} =0$.  
\end{enumerate}
The super-Weyl invariance of  ${\mathbb V}_{(n)} $ follows from 
the discussion in  \cite{CF}
(see also \cite{KMcC}).

Associated with ${\mathbb V}_{(n)} $ is the following generalised 
FI term \cite{K18}
\bea
{\mathbb J}^{(n)}_{\rm FI} [V; \U]=   \int \rd^4 x \rd^2 \q  \rd^2 \bar{\q} \, E\,
\U  {\mathbb V}_{(n)}~,
\label{FIterm}
\eea
where $\U$ is a real scalar with  super-Weyl transformation 
\bea
\d_\s \U = (\s +\bar \s) \U~.
\eea
The composite ${\mathbb V}_{(-1)}$  and 
the associated FI term ${\mathbb J}^{(-1)}_{\rm FI} $
were discovered in \cite{CFTVP}.

It is $\U$ which contains information about a specific off-shell supergravity theory.
Within the new minimal formulation for $\cN=1$ supergravity
\cite{new,SohniusW3}, $\U$ can be identified with the corresponding 
linear compensator\footnote{The linear compensator \cite{deWR}
is described by a tensor multiplet
\cite{Siegel-tensor} such that its field strength $L$ is
nowhere vanishing.} $ L$, 
\bea  
(\bar \cD^2 -4R) { L}  =0~, \qquad \bar { L}= { L}~.
\eea
In pure old minimal  supergravity
\cite{WZ,old1,old2}, $\U$ is given by 
$\U = \bar S_0 S_0$,
where  $S_0$ is 
the chiral compensator$, \bar \cD_\ad S_0 =0$,
with super-Weyl transformation law 
$\d_\s S_0 = \s S_0$. In the presence of chiral matter, however, $\U$ must be deformed, 
see below. It should be mentioned that the use of conformal compensators 
to describe off-shell formulations for supergravity 
was advocated  by many authors including 
\cite{SG,deWR,KU,FGKV}.

In the literature there have appeared various applications of the generalised FI terms
\cite{ACIK1,ACIK2,FKR,AKK,ST,AAAK1,NY1,NY2,CFT,AAAK2,A}, including inflationary cosmological models.

It is worth remarking that the right-hand side of \eqref{FIterm}
can be written in several equivalent forms using the identities
\bea
{\mathbb V} \,\bar \cD^2 \cD^2   
\frac{W^2 \bar W^2 }{(\cD W)^{4}} =  {\mathbb V} 
\frac{\cD^2 W^2 \bar \cD^2 \bar W^2 }{(\cD W)^{4}}
= {\mathbb V} \frac{\cD {\mathbb W} }{\cD W} ~,
\qquad {\mathbb W}_\a := -\frac{1}{4} (\bar \cD^2 - 4R) \cD_\a {\mathbb V}~.
\eea

Actually, it is possible to consider more general FI-like  terms of the form
\bea
{\mathbb J}^{(\cG)}_{\rm FI} [V; \U]=   \int \rd^4 x \rd^2 \q  \rd^2 \bar{\q} \, E\,
\U  {\mathbb V} \,\cG \left(- \frac{\cD^2 W^2  }{(\cD W)^{2}}
\,, \, -\frac{ \bar \cD^2 \bar W^2 }{(\cD W)^{2}} \right)~,
\label{generalised}
\eea
where $\cG (z, \bar z)$ is a real function of one complex variable. 
The component structure of ${\mathbb J}^{(\cG)}_{\rm FI} [V; \U]$ will be discussed below.
The original functional \eqref{FIterm} corresponds to the choice
$\cG (z, \bar z) =(z \bar z)^n$.

Each of the generalised FI terms \eqref{generalised}, 
including \eqref{FIterm}, is not superconformal in the sense that 
the integrand involves a conformal compensator. Quite remarkably, 
once the condition $\cD W\neq 0$ is emposed, 
it is also possible 
to construct superconformal self-couplings for the vector multiplet. 
Such superconformal self-couplings are proposed in this note. 
They are described by super-Weyl invariant functionals of the form
\bea
S [V] =
\frac{1}{2}  \int \rd^4 x \rd^2 \q   \, \cE\,
W^2 &+&
 \int \rd^4 x \rd^2 \q  \rd^2 \bar{\q} \, E\,
\frac{W^2\,{\bar W}^2}{(\cD W)^2}\,
\cH \left(- \frac{\cD^2 W^2  }{(\cD W)^{2}}
\,, \, -\frac{ \bar \cD^2 \bar W^2 }{(\cD W)^{2}} \right)~,
\label{action}
\eea
where $\cE$ is the chiral integration measure \cite{Siegel,Zumino}, 
and $\cH (z, \bar z)$ is a real function of one complex variable.
This action is part of the complete supergravity-matter action given by 
\bea
S= S_{\rm SUGRA} + S[V] -2\x {\mathbb J}^{(\cG)}_{\rm FI} [V; \U] 
\label{13}
\eea
where $S_{\rm SUGRA} $ denotes an action for supergravity coupled to other 
matter supermultiplets, for instance
\bea
S_{\rm SUGRA} &=& - 3
\int {\rm d}^{4} x \rd^2\q\rd^2\bar\q\,
E\,  \bar S_0  \,
\re^{-\frac{1}{3} K(\f , \bar \f)}S_0
+ \Big\{     \int {\rm d}^{4} x \rd^2 \q\,
{\cal E} \, S_0^3   W(\f)
+ {\rm c.c.} \Big\}
\eea
which corresponds to the old minimal formulation for $\cN=1$ supergravity. 
In this case $\U$ in the generalised FI term in \eqref{13}
should be $\U = \bar S_0  \re^{-\frac{1}{3} K(\f , \bar \f)}S_0 $, as was pointed out 
in \cite{ACIK2,AKK}.

In general, the action \eqref{action} is highly nonlinear. 
However its functional form drastically simplifies provided 
the ordinary gauge field contained in $V$ is chosen to be a flat connection.
This means that the gauge freedom \eqref{gaugefreedom}
may be used to make $V $ a nilpotent superfield
obeying the constraints 
\bea
{V}{V}=0~, 
\qquad
{V} \cD_A \cD_B {V}=0~,
\qquad
{ V} \cD_A \cD_B \cD_C {V} =0~.
\label{14}
\eea
Then it can be seen that 
\bea
{\mathbb V}_{(n)} =V~,
\eea
compare with the analysis in \cite{KMcAT-M}. This implies 
\bea
S[V] -2f {\mathbb J}^{(n)}_{\rm FI} [V; \U] 
&=& \frac{h}{2}  \int \rd^4 x \rd^2 \q   \, \cE\,   
W^2 -2\x g \int \rd^4 x \rd^2 \q  \rd^2 \bar{\q} \, E\,
\U V~, 
\label{Goldstino}
\eea
where we have denoted 
\bea
h:=1 + \hf \cH (1, 1)>0~, \qquad g := \cG(1,1)\neq0~.
\label{inequality}
\eea
Modulo an overall numerical factor, \eqref{Goldstino} is the  Goldstino multiplet action 
proposed in \cite{KMcAT-M}. The restriction $h>0$
is equivalent to the requirement that the Goldstino kinetic term  has
 the correct sign.  At the component level, the action \eqref{Goldstino}
is still nonlinear, due to the nilpotency constraints  \eqref{14}.
However, the functional form of the action \eqref{Goldstino} is universal, 
unlike the complete vector multiplet action in \eqref{13}, which is a manifestation 
of the universality of the Volkov-Akulov action \cite{VA,IK}.\footnote{All the constraints 
\eqref{14} are invariant under local rescalings $V \to \re^{\r} V$, with the parameter $\r$ being an arbitrary real scalar superfield. Requiring the action \eqref{Goldstino} to be 
stationary under such rescalings gives the constraint 
$f \U V =\frac{1}{8} V \cD^\a (\bar \cD^2 -4R ) \cD_\a V$, where $f=\x g/h$. 
In conjunction with \eqref{14},  this constraint defines
the irreducible Goldstino multiplet introduced in \cite{BHKMS}.}

To arrive at the Goldstino multiplet action \eqref{Goldstino}, we have made use 
of the gauge \eqref{14} which expresses the fact that the gauge field is switched off.
 It is actually possible to avoid imposing any gauge condition. 
In general the gauge prepotential $V$ may be split in two multiplets, one of which 
contains only a single independent component field -- the gauge field itself -- while
the second part contains the remaining component fields.
The point is that the nilpotency conditions \eqref{5} allow us to interpret ${\mathbb V}_{(n)}$
as a Goldstino superfield of the type proposed in \cite{KMcAT-M} provided
 its $D$-field is nowhere vanishing, which means that $\cD^2 {W}^2$ is nowhere vanishing, 
 in addition to the condition $\cD W\neq 0$ imposed earlier. 
Then ${\mathbb V}_{(n)}$ contains only two independent component fields, 
the Goldstino and $D$-field. We then can introduce a  new parametrisation 
for the gauge prepotential given by 
\bea
V = A_{(n)} + {\mathbb V}_{(n)}~.
\eea
 It is $A_{(n)} $ which varies under the gauge transformation \eqref{gaugefreedom},
$\d_\l A_{(n)}  = \l + \bar \l$, while ${\mathbb V}_{(n)}$ is gauge invariant by construction.
Modulo purely gauge  degrees of freedom, $A_{(n)} $ contains only one independent 
field, the gauge field.

It is of interest to work out the bosonic sector of the model \eqref{13}
in the vector multiplet sector. For this purpose 
we  introduce  gauge-invariant component fields of the vector multiplet following \cite{KMcC}
\bea
W_{\a}\arrowvert = \j_{\a}~,\qquad
-\frac{1}{2}\cD^{\a}W_{\a}\arrowvert=D~,
\qquad
\cD_{(\a}W_{\b)}\arrowvert 
=2 {\rm i} {\hat F}_{\a\b}
&=& {\rm i} (\s^{ab})_{\a\b}{\hat F}_{ab}~,
\label{19}
\eea
where the bar-projection  $U|$ of a superfield $U$ means switching off the superspace Grassmann variables, and 
\bea
\label{eq:F_ab defn}
{\hat F}_{ab} &=& F_{ab} -
\frac{1}{2}(\J_{a}\s_{b}{\bar \j} + 
\j\s_{b}{\bar \J}_{a}) +
\frac{1}{2}(\J_{b}\s_{a}{\bar \j} + 
\j\s_{a}{\bar \J}_{b})~,
\non\\
F_{ab} &=& {\bm \nabla}_{a}V_{b} 
- {\bm \nabla}_{b}V_{a} 
- {\cT_{ab}}^{c}V_{c}~,
\eea
with $V_a= e_a{}^m (x) \,V_m (x)$  
the  gauge  one-form, and $\J_a{}^\b$ the gravitino. 
Here ${\bm \nabla}_a$ denotes a spacetime covariant derivative with torsion,
\bea
\label{eq:covariant derivative algebra}
\left[{\bm \nabla}_{a}, {\bm \nabla}_{b}\right] = {\cT_{ab}}^{c} {\bm \nabla}_{c}
+ \frac{1}{2}\,\cR_{abcd} M^{cd}~,
\eea
where $\cR_{abcd}$ is the curvature tensor and $\cT_{abc}$ 
is the torsion tensor. The latter is related to the gravitino by
\bea
\cT_{abc} = -\frac{\rm i}{2}(\J_{a}\s_{c}{\bar \J}_{b}
- \J_{b}\s_{c}{\bar \J}_{a})~.
\eea
For more details, see \cite{Ideas,KMcC}.
We deduce from the above relations that 
\bea
-\frac{1}{4} \cD^2 {W}^2| = D^2 - 2 F^2 +
\text{fermionic terms}~, \qquad F^2: =F^{\a\b} F_{\a\b}~.
\eea
We conclude that the electromagnetic field should be weak enough 
to satisfy $D^2 - 2 F^2  \neq 0$,
in addition to the condition $D \neq 0$ discussed above.
Direct calculations give the component bosonic Lagrangian 
\bea
\cL (F_{ab} , D)= -\hf (F^2 +\bar F^2) &+&\hf D^2\left\{
1+ \hf  \cH \Big(1 -  \frac{2F^2}{D^2}\, ,\, 1 -  \frac{2\bar F^2}{D^2} \Big) 
\Big| 1 -  \frac{2F^2}{D^2} \Big|^2
 \right\} \non \\
&-&\x  D\, 
\cG \Big(1 -  \frac{2F^2}{D^2}\, ,\, 1 -  \frac{2\bar F^2}{D^2} \Big) 
\Big| 1 -  \frac{2F^2}{D^2} \Big|^2 \U|~.
\label{23}
\eea
In order for the supergravity action in \eqref{13} to give  the correct Einstein-Hilbert gravitational Lagrangian at
the component level, one has to impose the super-Weyl gauge
$\U|=1$, see \cite{KU,KMcC} for the technical details.

It is seen that the case \cite{CFTVP} 
\bea
\cG (z, \bar z) =(z \bar z)^{-1}
\label{unique}
\eea
is special since the last term becomes linear in $D$ and independent of the field strength. This simplicity is somewhat misleading since 
the generalised FI term \eqref{FIterm}
is nonlinear in the Goldstino for arbitrary $n$. 
Of course, this nonlinear fermionic sector disappears in the unitary gauge
in which the Goldstino is gauged away.
However, we have shown that the model \eqref{13} describes spontaneously broken local supersymmetry for any choice of $\cG$, and hence
 there is nothing unique in the choice \eqref{unique}
from the conceptual point of view.

As follows from \eqref{23}, 
the auxiliary field $D$ may be integrated out (at least in perturbation theory) using its equation of motion 
\bea
\frac{\pa}{\pa D} \cL (F_{ab} , D)=0~,
\eea
leaving a model for nonlinear electrodynamics, ${\mathfrak L} (F_{ab})$.

Action \eqref{action} is superconformal since it describes the self-interacting 
vector multiplet coupled to conformal supergravity. In the presence of a conformal compensator, which corresponds to an off-shell supergravity theory, 
more general couplings exist. 
Ref.\,\cite{KMcC} presented a general family of $\sU(1)$ duality invariant models
for a massless vector multiplet coupled to off-shell supergravity, 
old minimal or new minimal.
Such a theory is described by a super-Weyl invariant action of the form
\bea
S_{\text{self-dual}}[V;\U] =
\frac{1}{2}  \int \rd^4 x \rd^2 \q   \, \cE\,
W^2 &+&
\frac14  \int \rd^4 x \rd^2 \q  \rd^2 \bar{\q} \, E\,
\frac{W^2\,{\bar W}^2}{\U^2}\,
\L\!\left(\frac{\o}{\U^2},
\frac{\bar \o}{\U^2}\right)~.
\label{25}
\eea
Here 
$\o:=  \frac{1}{8} \cD^2 W^2$, and $\L(\o,\bar \o)$ is a real analytic 
function satisfying the equation \cite{KT1,KT2}
\bea
{\rm Im} \,\Big\{ \G
- \bar{\o}\, \G^2
\Big\} = 0~, \qquad \quad
\G  := \frac{\pa (\o \, \L) }{\pa \o}~.
\label{26}
\eea
These $\sU(1)$ duality invariant  theories 
are curved-superspace extensions 
of the globally supersymmetric systems 
introduced in \cite{KT1,KT2}.
The supersymmetric Born-Infeld action
coupled to supergravity  \cite{CF} is obtained by choosing
\bea
&&\L_{\rm SBI}(\o,{\bar\o}) =
\frac{k^2}
{ 1 + \hf\, A \, + \sqrt{1 + A +\frac{1}{4} \,B^2} }~,
\quad
A=k^2(\o  + \bar \o)~, \quad 
B=k^2(\o - \bar \o)~,
\label{27}
\eea
with $k$ a coupling constant.
It was pointed out by Cecotti and Ferrara \cite{CF} that the dynamical system defined by 
eqs.\,\eqref{25} and \eqref{27} is not a unique supersymmetric extension of the 
Born-Infeld action. One can introduce a two-parameter deformation of \eqref{27} 
obtained by replacing 
\bea
\o \to \o +\z \cD W~, \qquad \z ={\rm const}~,
\label{28}
\eea
 with $\z$ a complex parameter.  
At the component level, the resulting bosonic action coincides with the Born-Infeld one
provided the auxiliary field is switched off. However, the freedom to perform shifts \eqref{28} 
is eliminated if one requires the supersymmetric theory to possess $\sU(1)$ duality 
invariance. In other words, no $\cD W $-dependence is allowed in duality invariant 
models. The same condition emerges if the $\cN=1$ vector multiplet 
is used to describe partial $\cN=2 \to \cN=1$ supersymmetry
breaking \cite{BG}.

As pointed out in \cite{ADM},
an important property of the standard FI term
is that it remains invariant under the second nonlinearly 
realised supersymmetry of the rigid supersymmetric Born-Infeld 
action \cite{BG}. This property implies the supersymmetric Born-Infeld 
action deformed by a FI term still describes partial $\cN=2 \to \cN=1$ supersymmetry
breaking \cite{ADM,K-FI,DFS}, and the resulting model is compatible with $\sU(1)$ 
duality invariance \cite{K-FI}.
As for the generalised FI-type terms \eqref{FIterm}, 
they do not  share these fundamental properties. 

One may compare \eqref{action} with general $\cN=2$ superconformal 
actions for a vector multiplet 
in Minkowski superspace \cite{BKT}
\bea
\G =  \int \rd^4 x \rd^4 \q  \rd^4 \bar{\q} \, \Big\{ \frac{c}{2}  \ln \cW \ln \bar \cW
&+& \ln \cW \L(\bar \cW^{-2} \ln \cW) +{\rm c.c.}  \non \\
&+& \S ( \bar \cW^{-2} D^4 \ln \cW , \cW^{-2} \bar \cD^4 \ln \bar \cW ) \Big\}~,
\label{29}
\eea
where $\cW$ is a reduced $\cN=2$ chiral superfield constrained by 
\bea
\bar D^\ad_i \cW=0~, \qquad D^{ij} \cW = \bar D^{ij} \bar \cW ~, \qquad 
D^{ij} := D^{\a(i}D^{j)}_\a ~, \quad \bar D^{ij} := \bar D_\ad^{(i} \bar D^{j) \,\ad} ~,
\label{30}
\eea
which describes the field strength of the vector multiplet.\footnote{The action 
\eqref{29} and constraints \eqref{30} involve $\cN=2$ spinor covariant derivatives $D_\a^i$ and $\bar D^\ad_i $, 
with $i =\1, \2$, and the fourth-order operators $D^4$ and $\bar D^4$ are defined as
$D^4 = (D^\1)^2 (D^\2)^2$ and $\bar D^4 = (\bar D_\1)^2 (\bar D_\2)^2$.}
It is assumed in \eqref{29} that the physical complex scalar of the vector multiplet, 
$\cW|$, is nowhere vanishing. Unlike the $\cN=1$ case considered in this paper, 
no assumption is made about the auxiliary iso-triplet, $D^{ij} \cW|$, since the case of unbroken $\cN=2$ supersymmetry is studied. Because of unbroken supersymmetry, 
all contributions containing factors of the {\it primary} superfield $D^{ij} \cW$ are omitted due to the fact that this primary operator constitutes the free equation of motion (the functional \eqref{29} is interpreted as a low-energy effective action). 
The $\cN=1$ analogue of 
 $D^{ij} \cW$ is the super-Weyl primary multiplet $\cD W$ we have used above. 
 Since in the $\cN=1$ case we are interested in models for spontaneously broken supersymmetry, which means $\cD W$ is nowhere vanishing,
 we are no longer allowed to discard terms involving factors of $\cD W$. 
 It is for these reasons that  actions of the form \eqref{action} have to be considered. 

Let us summarise the main results of this paper. We proposed 
the new generalised FI terms \eqref{generalised} which include those constructed earlier \cite{K18,CFTVP}. 
We  introduced  new models for spontaneously broken local supersymmetry
\eqref{13} which make use of the novel superconformal 
vector multiplet self-couplings \eqref{action}.

The constructions given in this paper have a natural extension to $\cN = 2$ supergravity, which will be described elsewhere.
\\

%%%%%%%%%%%%%%%%%%%%%%%%%%%%%%%%%%%%%%%%%%%%%%%%%%%%%%
%%%%%%%%%%%%%%%%%%%%%%%%%%%%%%%%%%%%%%%%%%%%%%%%%%%%%%

\noindent
{\bf Acknowledgements:}  It is my pleasure to thank Tsutomu Yanagida for
hospitality at the Kavli Institute for the Physics and Mathematics of the Universe, Tokyo,
where this project was initiated.  I am grateful to Darren Grasso and Dmitri Sorokin
for constructive comments on the manuscript. 
This  work is supported in part by the Australian 
Research Council, project No. DP160103633.

\appendix 

\numberwithin{equation}{section}

\section{Conformal supergravity in superspace}

In the framework of the vielbein formulation for conformal gravity,
the gauge field is  a vielbein $e^a := \rd x^m e_m{}^a (x)$, 
  $e:=\det (e_m{}^a ) \neq 0$,  while the metric becomes a composite field 
  defined by $g_{mn} = e_m{}^a e_n{}^b \eta_{ab} $, 
with $\eta_{ab}$ the Minkowski metric. 
The gauge group of conformal gravity is spanned by  general coordinate, local Lorentz 
and Weyl transformations which act on the torsion-free covariant derivatives
\bea
\nabla_a = e_a +\o_a = e_a{}^m   \pa_m +\hf \o_a{}^{bc}  M_{bc} ~, \qquad 
[\nabla_a , \nabla_b ] = \hf R_{ab}{}^{cd} M_{cd} 
\label{A.1}
\eea
by the rule
\bea
\d \nabla_a &=&  [ \x^b \nabla_b +\hf K^{bc}M_{bc} , \nabla_a]
+  \t \nabla_a +(\nabla^b\t) M_{ba} ~,
\eea
with the gauge parameters $\x^a (x) = \x^m (x) e_m{}^a (x) $,
$K^{ab} (x) = - K^{ba}(x)$ and $\t(x)$
being completely arbitrary.
In  \eqref{A.1}, $M_{bc} = -M_{cb}$ is the Lorentz generator, 
$ e_a{}^m (x) $ the inverse vielbein, $e_a{}^m e_m{}^b = \d_a{}^b$,  
and $\o_a{}^{bc} (x)$ the torsion-free Lorentz connection.

In order to describe $\cN=1$ conformal supergravity \cite{KTvN1,KTvN2}
in superspace, the simplest approach is to make use 
of the Grimm-Wess-Zumino geometry \cite{GWZ}, which is at the heart of
the Wess-Zumino formulation for old minimal supergravity \cite{WZ}, 
in conjunction with the super-Weyl transformations discovered  
by Howe and Tucker  \cite{HT}.
 The geometry of curved superspace is described by covariant derivatives
of the form
\bea
\cD_{A}=(\cD_{{a}}, \cD_{{\a}},\cDB^\ad)
=E_{A}{}^M \pa_M
+\hf \O_{{A}}{}^{bc}M_{bc} 
~,
\label{CovDev}
\eea
which obey the graded commutation relations 
(see \cite{Ideas} for a derivation)
\begin{subequations}\label{algebra}
\bea
& \{ \cD_\a , {\bar \cD}_\ad \} = -2{\rm i} \cD_{\a \ad} ~,\\
&
\{\cD_\a, \cD_\b \} = -4{\bar R} M_{\a \b}~,
 \qquad
\{ {\bar \cD}_\ad, {\bar \cD}_\bd \} =  4R {\bar M}_{\ad \bd}~, 
 \\
&\left[ \cD_{\a} , \cD_{ \b \bd } \right]
      = 
     {\rm i}  {\ve}_{\a \b}
\Big({\bar R}\,\cDB_\bd + G^\g{}_\bd \cD_\g
- \cD^\g G^\d{}_\bd  M_{\g \d}
+2{\bar W}_\bd{}^{\gd \dot{\d}}
{\bar M}_{\gd \dot{\d} }  \Big)
+ {\rm i} \cDB_\bd {\bar R} \, M_{\a \b}~,
~~~~~~\\
&\left[ { \bar \cD}_{\ad} , \cD_{ \b \bd } \right]
      =  -{\rm i}{\ve}_{\ad \bd}
\Big(R\,\cD_\b + G_\b{}^{\dot{\g}}  \cDB_{\dot{\g}}
-\cDB^\gd G_\b{}^{\dot{\d}}
{\bar M}_{\gd \dot{\d}}
+2W_\b{}^{\g \d}
M_{\g \d} \Big)
- {\rm i} \cD_\b R  {\bar M}_{\ad \bd}~.~~~~~~~ ~~
\eea
\esubeq
Here the torsion tensors $R$, $G_a = {\bar G}_a$ and
$W_{\a \b \g} = W_{(\a \b\g)}$ satisfy the  Bianchi identities:
\begin{subequations}
\bea
&\cDB_\ad R= 0~,~~~~~~\cDB_\ad W_{\a \b \g} = 0~,
\\
&
\cDB^\gd G_{\a \gd} = \cD_\a R~,~~~~~~
\cD^\g W_{\a \b \g} = {\rm i} \,\cD_{(\a }{}^\gd G_{\b) \gd}~.
\eea
\end{subequations}
The super-Weyl transformations are
\begin{subequations} 
\label{superweyl}
\bea
\d_\s \cD_\a &=& ( {\bar \s} - \hf \s)  \cD_\a + \cD^\b \s \, M_{\a \b}  ~, \\
\d_\s \bar \cD_\ad & = & (  \s -  \hf {\bar \s})
\bar \cD_\ad +  ( \bar \cD^\bd  {\bar \s} )  {\bar M}_{\ad \bd} ~,\\
\d_\s \cD_{\a\ad} &=& \hf( \s +\bar \s) \cD_{\a\ad} 
+\frac{\ri}{2} \bar \cD_\ad \bar \s \,\cD_\a + \frac{\ri}{2}  \cD_\a  \s\, \bar \cD_\ad
+ \cD^\b{}_\ad \s\, M_{\a\b} + \cD_\a{}^\bd \bar \s\, \bar M_{\ad \bd}~,
~~~~~~
\eea
\end{subequations}
accompanied by 
the following transformations of the torsion superfields
\begin{subequations} 
\bea
\d_\s R &=& 2\s R +\frac{1}{4} (\bar \cD^2 -4R ) \bar \s ~, \\
\d_\s G_{\a\ad} &=& \hf (\s +\bar \s) G_{\a\ad} +\ri \cD_{\a\ad} ( \s- \bar \s) ~, 
\label{s-WeylG}\\
\d_\s W_{\a\b\g} &=&\frac{3}{2} \s W_{\a\b\g}~.
\label{s-WeylW}
\eea
\end{subequations} 
Here the super-Weyl parameter $\s$ is a covariantly chiral scalar superfield,  $\bar \cD_\ad \s =0$. The super-Weyl transformations belong to the  
gauge group of conformal supergravity.

A tensor superfield $\mathfrak T $ (with its  indices suppressed)
is said to be super-Weyl primary of weight $(p,q)$
if its super-Weyl transformation law is 
$ \d_\s {\mathfrak T} =\big(p\, \s + q\, \bar \s \big) {\mathfrak T}$,
for some parameters $p$ and $q$.

There exist two other superspace formulations for conformal supergravity
\cite{Howe,Butter} which have structure groups larger than $\sSL (2,{\mathbb C})$.
These formulations are not used in the present paper, although our results 
can readily be lifted to $\sU(1)$ superspace \cite{Howe} and conformal superspace
\cite{Butter}. At the component level, the latter approach naturally reduces to the superconformal tensor calculus \cite{KTvN2,KakuT,FGvN,TvN} 
as demonstrated in \cite{Butter,KYY1,KYY2}.

%%%%%%%%%%%%%%%%%%%%%%%%%%%%%%%%%%%%%%%%%%%%%%%%%%%%%%
%%%%%%%%%%%%%%%%%%%%%%%%%%%%%%%%%%%%%%%%%%%%%%%%%%%%%%

\begin{footnotesize}

\end{footnotesize}


\begin{thebibliography}{66}

\bibitem{K18} 
  S.~M.~Kuzenko,
  ``Taking a vector supermultiplet apart: Alternative Fayet-Iliopoulos-type terms,''
  Phys. Lett. B {\bf 781}, 723 (2018)
%  doi:10.1016/j.physletb.2018.04.051
  [arXiv:1801.04794 [hep-th]].

\bibitem{CFTVP} 
  N.~Cribiori, F.~Farakos, M.~Tournoy and A.~Van Proeyen,
  ``Fayet-Iliopoulos terms in supergravity without gauged R-symmetry,''
JHEP {\bf 1804}, 032 (2018)
%  doi:10.1007/JHEP04(2018)032
  [arXiv:1712.08601 [hep-th]].

%\cite{Fayet:1974jb}
\bibitem{FI}
P.~Fayet and J.~Iliopoulos,
``Spontaneously broken supergauge symmetries and Goldstone spinors,''
 Phys.\ Lett.\  B {\bf 51}, 461 (1974).
  %%CITATION = PHLTA,B51,461;%%

\bibitem{Freedman} 
  D.~Z.~Freedman,
  ``Supergravity with axial gauge invariance,''
  Phys.\ Rev.\ D {\bf 15}, 1173 (1977).
  
\bibitem{BFNS} 
  R.~Barbieri, S.~Ferrara, D.~V.~Nanopoulos and K.~S.~Stelle,
  ``Supergravity, R invariance and spontaneous supersymmetry breaking,''
  Phys.\ Lett.\  {\bf 113B}, 219 (1982).  

\bibitem{FZ} 
  S.~Ferrara and B.~Zumino,
  ``Supergauge invariant Yang-Mills theories,''
  Nucl.\ Phys.\ B {\bf 79}, 413 (1974).

\bibitem{WZ}
J.~Wess and B.~Zumino,
 ``Superfield Lagrangian for supergravity,''
 Phys.\ Lett.\  B {\bf 74}, 51 (1978).
%%CITATION = PHLTA,B74,51;%%


 \bibitem{Ideas} 
I.~L. Buchbinder and S.~M. Kuzenko, {\it Ideas and Methods of Supersymmetry and
Supergravity, Or a Walk Through Superspace},
 IOP, Bristol, 1995 (Revised Edition 1998).



\bibitem{KMcAT-M} 
  S.~M.~Kuzenko, I.~N.~McArthur and G.~Tartaglino-Mazzucchelli,
  ``Goldstino superfields in N=2 supergravity,''   JHEP {\bf 1705}, 061 (2017)
[arXiv:1702.02423 [hep-th]].


\bibitem{CF}
S.~Cecotti and S.~Ferrara,
``Supersymmetric Born-Infeld Lagrangians,''
Phys.\ Lett.\ B {\bf 187}, 335 (1987).


%\cite{Kuzenko:2005wh}
\bibitem{KMcC} 
  S.~M.~Kuzenko and S.~A.~McCarthy,
  ``On the component structure of N=1 supersymmetric nonlinear electrodynamics,''
  JHEP {\bf 0505}, 012 (2005)
  %doi:10.1088/1126-6708/2005/05/012
  [hep-th/0501172].



\bibitem{new}
M.~F.~Sohnius and P.~C.~West,
``An alternative minimal off-shell version of N=1 supergravity,''
Phys.\ Lett.\  B {\bf 105}, 353 (1981).
  
    
\bibitem{SohniusW3} 
 M.~Sohnius and P.~C.~West,
 ``The tensor calculus and matter coupling of the alternative minimal auxiliary field formulation of $N=1$ supergravity,''  Nucl.\ Phys.\ B {\bf 198}, 493 (1982).    


\bibitem{deWR}
B.~de Wit and M.~Ro\v{c}ek,
``Improved tensor multiplets,''
Phys.\ Lett.\ B {\bf 109}, 439 (1982).

\bibitem{Siegel-tensor}
W.~Siegel,
``Gauge spinor superfield as a scalar multiplet,''
Phys.\ Lett.\ B {\bf 85}, 333 (1979).

\bibitem{old1}
K.~S.~Stelle and P.~C.~West,
``Minimal auxiliary fields for supergravity,''
Phys.\ Lett.\  B {\bf 74},  330 (1978).

\bibitem{old2}
S.~Ferrara and P.~van Nieuwenhuizen,
``The auxiliary fields of supergravity,''
Phys.\ Lett.\  B {\bf 74}, 333 (1978).



\bibitem{SG}
W.~Siegel and S.~J.~Gates Jr.
 ``Superfield supergravity,''  Nucl.\ Phys.\  B {\bf 147}, 77 (1979).
  

\bibitem{KU}
T.~Kugo and S.~Uehara,
``Improved superconformal gauge conditions in 
the N=1 supergravity  Yang-Mills matter system,''
Nucl.\ Phys.\ B {\bf 222}, 125 (1983).



%\cite{Ferrara:1983dh}
\bibitem{FGKV}
 S.~Ferrara, L.~Girardello, T.~Kugo and A.~Van Proeyen,
``Relation between different auxiliary field formulations of N=1 supergravity
coupled to matter,''   Nucl.\ Phys.\  B {\bf 223}, 191 (1983).
  %%CITATION = NUPHA,B223,191;%

\bibitem{ACIK1} 
  I.~Antoniadis, A.~Chatrabhuti, H.~Isono and R.~Knoops,
  ``Fayet-Iliopoulos terms in supergravity and D-term inflation,''
  Eur.\ Phys.\ J.\ C {\bf 78}, no. 5, 366 (2018)
  %doi:10.1140/epjc/s10052-018-5861-6
  [arXiv:1803.03817 [hep-th]].

\bibitem{ACIK2} 
  I.~Antoniadis, A.~Chatrabhuti, H.~Isono and R.~Knoops,
  ``The cosmological constant in zupergravity,''
  Eur.\ Phys.\ J.\ C {\bf 78}, no. 9, 718 (2018)
%  doi:10.1140/epjc/s10052-018-6175-4
  [arXiv:1805.00852 [hep-th]].

\bibitem{FKR} 
  F.~Farakos, A.~Kehagias and A.~Riotto,
  ``Liberated $ \mathcal{N} $ = 1 supergravity,''
  JHEP {\bf 1806}, 011 (2018)
  %doi:10.1007/JHEP06(2018)011
  [arXiv:1805.01877 [hep-th]].
  
\bibitem{AKK} 
  Y.~Aldabergenov, S.~V.~Ketov and R.~Knoops,
  ``General couplings of a vector multiplet in $N=1$ supergravity with new FI terms,''
  Phys.\ Lett.\ B {\bf 785}, 284 (2018)
  %doi:10.1016/j.physletb.2018.07.072
  [arXiv:1806.04290 [hep-th]].  
  
   
 \bibitem{ST} 
  K.~Schmitz and T.~T.~Yanagida,
  ``Axion isocurvature perturbations in low-scale models of hybrid inflation,''
  Phys.\ Rev.\ D {\bf 98}, no. 7, 075003 (2018)
 % doi:10.1103/PhysRevD.98.075003
  [arXiv:1806.06056 [hep-ph]]. 

  
\bibitem{AAAK1} 
  H.~Abe, Y.~Aldabergenov, S.~Aoki and S.~V.~Ketov,
  ``Massive vector multiplet with Dirac-Born-Infeld and new Fayet-Iliopoulos terms in supergravity,''
  JHEP {\bf 1809}, 094 (2018)
  %doi:10.1007/JHEP09(2018)094
  [arXiv:1808.00669 [hep-th]].  
   
 \bibitem{NY1} 
  M.~Nitta and R.~Yokokura,
  ``Higher derivative three-form gauge theories and their supersymmetric extension,''
  JHEP {\bf 1810}, 146 (2018)
%  doi:10.1007/JHEP10(2018)146
  [arXiv:1809.03957 [hep-th]]. 
  
 \bibitem{NY2} 
  M.~Nitta and R.~Yokokura,
  ``Topological couplings in higher derivative extensions of supersymmetric three-form gauge theories,''
  arXiv:1810.12678 [hep-th].
  %%CITATION = ARXIV:1810.12678;%% 
  
\bibitem{CFT} 
  N.~Cribiori, F.~Farakos and M.~Tournoy,
  ``Supersymmetric Born-Infeld actions and new Fayet-Iliopoulos terms,''
  JHEP {\bf 1903}, 050 (2019)
  %doi:10.1007/JHEP03(2019)050
  [arXiv:1811.08424 [hep-th]].  

\bibitem{AAAK2} 
  H.~Abe, Y.~Aldabergenov, S.~Aoki and S.~V.~Ketov,
  ``Polonyi-Starobinsky supergravity with inflation in a massive vector multiplet with DBI and FI terms,''
  Class.\ Quant.\ Grav.\  {\bf 36}, no. 7, 075012 (2019)
  %doi:10.1088/1361-6382/ab0901
  [arXiv:1812.01297 [hep-th]].
  
\bibitem{A} 
  Y.~Aldabergenov,
  ``No-scale supergravity with new Fayet-Iliopoulos term,''
  arXiv:1903.11829 [hep-th].  
  

\bibitem{Siegel}
W.~Siegel,
``Solution to constraints in Wess-Zumino supergravity formalism,''
Nucl.\ Phys.\  B {\bf 142}, 301 (1978). 


\bibitem{Zumino} 
  B.~Zumino,
  ``Supergravity and superspace,''
in {\it Recent Developments in  Gravitation - Carg\`ese 1978}, 
M. L\'evy and S. Deser (Eds.), N.Y., Plenum Press, 1979, pp. 405--459.


\bibitem{VA}
D.~V.~Volkov and V.~P.~Akulov,
``Possible universal neutrino interaction,''
  {JETP Lett.\  {\bf 16}, 438 (1972)}   
  [Pisma Zh.\ Eksp.\ Teor.\ Fiz.\   {\bf 16},  621 (1972)]; 
  ``Is the neutrino a Goldstone particle?,''
  Phys.\ Lett.\  B {\bf 46}, 109 (1973).

\bibitem{IK}
E.~Ivanov and A.~Kapustnikov, ``General relationship between linear and
  nonlinear realisations of supersymmetry,''
  J.\ Phys.\  A {\bfseries 11},  2375  (1978);
``The nonlinear realisation structure of models
  with spontaneously broken supersymmetry,''
   J.\ Phys.\ G  {\bf 8},  167 (1982). 



\bibitem{BHKMS} 
  I.~Bandos, M.~Heller, S.~M.~Kuzenko, L.~Martucci and D.~Sorokin,
  ``The Goldstino brane, the constrained superfields and matter in $ \mathcal{N}=1 $ supergravity,''  
  JHEP {\bf 1611}, 109 (2016)
  [arXiv:1608.05908 [hep-th]].
  


\bibitem{KT1}
S.~M.~Kuzenko and S.~Theisen,
``Supersymmetric duality rotations,''
JHEP {\bf 0003}, 034 (2000)
[arXiv:hep-th/0001068].



\bibitem{KT2}
S.~M.~Kuzenko and S.~Theisen,
``Nonlinear self-duality and supersymmetry,''
Fortsch.\ Phys.\  {\bf 49}, 273 (2001)
[arXiv:hep-th/0007231].


\bibitem{BG}
J.~Bagger and A.~Galperin,
``A new Goldstone multiplet for partially broken supersymmetry,''
Phys.\ Rev.\ D {\bf 55}, 1091 (1997) [arXiv:hep-th/9608177].


  
%\cite{Antoniadis:2008uk}
\bibitem{ADM}
I.~Antoniadis, J.~P.~Derendinger and T.~Maillard,
``Nonlinear N=2 supersymmetry, effective actions and moduli stabilization,''
Nucl.\ Phys.\  B {\bf 808}, 53 (2009)  [arXiv:0804.1738 [hep-th]].
  %%CITATION = NUPHA,B808,53;%%

%\cite{Kuzenko:2009ym}
\bibitem{K-FI}
S.~M.~Kuzenko,
``The Fayet-Iliopoulos term and nonlinear self-duality,''
 Phys.\ Rev.\  D {\bf 81}, 085036 (2010)
 [arXiv:0911.5190 [hep-th]].
  %%CITATION = PHRVA,D81,085036;%%
  
\bibitem{DFS} 
E.~Dudas, S.~Ferrara and A.~Sagnotti,
 ``A superfield constraint for $ \mathcal{N}  = 2 \to \mathcal{N} $ = 0 breaking,''
  JHEP {\bf 1708}, 109 (2017)
  [arXiv:1707.03414 [hep-th]].



\bibitem{BKT} 
  I.~L.~Buchbinder, S.~M.~Kuzenko and A.~A.~Tseytlin,
  ``On low-energy effective actions in N=2, N=4 superconformal theories in four-dimensions,''
  Phys.\ Rev.\ D {\bf 62}, 045001 (2000)
%  doi:10.1103/PhysRevD.62.045001
  [hep-th/9911221].



\bibitem{KTvN1} 
  M.~Kaku, P.~K.~Townsend and P.~van Nieuwenhuizen,
  ``Gauge theory of the conformal and superconformal group,''
  Phys.\ Lett.\  {\bf 69B}, 304 (1977).

\bibitem{KTvN2} 
  M.~Kaku, P.~K.~Townsend and P.~van Nieuwenhuizen,
  ``Properties of conformal supergravity,''
  Phys.\ Rev.\ D {\bf 17}, 3179 (1978).



\bibitem{GWZ} 
  R.~Grimm, J.~Wess and B.~Zumino,
  ``Consistency checks on the superspace formulation of supergravity,''
  Phys.\ Lett.\ B {\bf 73}, 415 (1978);
  ``A complete solution of the Bianchi identities in superspace,''
 Nucl.\ Phys.\ B {\bf 152}, 255 (1979).


\bibitem{HT}
P.~S.~Howe and R.~W.~Tucker,
``Scale invariance in superspace,''
Phys.\ Lett.\ B {\bf 80}, 138 (1978).


\bibitem{Howe}
P.~S.~Howe,
``A superspace approach to extended conformal supergravity,''
  Phys.\ Lett.\ B {\bf 100}, 389 (1981);
``Supergravity in superspace,''  Nucl.\ Phys.\  B {\bf 199}, 309 (1982).
  
%\cite{Butter:2009cp}
\bibitem{Butter}
D.~Butter,  ``N=1 conformal superspace in four dimensions,''
Annals Phys.\  {\bf 325}, 1026 (2010)
  [arXiv:0906.4399 [hep-th]].
  %%CITATION = APNYA,325,1026;%%

\bibitem{KakuT} 
  M.~Kaku and P.~K.~Townsend,
  ``Poincar\'e supergravity as broken superconformal gravity,''
  Phys.\ Lett.\  {\bf 76B}, 54 (1978).
  %doi:10.1016/0370-2693(78)90098-9
  
\bibitem{FGvN} 
  S.~Ferrara, M.~T.~Grisaru and P.~van Nieuwenhuizen,
  ``Poincar\'e and conformal supergravity models with closed algebras,''
  Nucl.\ Phys.\ B {\bf 138}, 430 (1978).  
  
  \bibitem{TvN} 
  P.~K.~Townsend and P.~van Nieuwenhuizen,
  ``Simplifications of conformal supergravity,''
  Phys.\ Rev.\ D {\bf 19}, 3166 (1979).
  
 \bibitem{KYY1} 
  T.~Kugo, R.~Yokokura and K.~Yoshioka,
  ``Component versus superspace approaches to D = 4, N = 1 conformal supergravity,''
  PTEP {\bf 2016}, no. 7, 073B07 (2016)
 % doi:10.1093/ptep/ptw090
  [arXiv:1602.04441 [hep-th]]. 
  
\bibitem{KYY2} 
  T.~Kugo, R.~Yokokura and K.~Yoshioka,
  ``Superspace gauge fixing in Yang-Mills matter-coupled conformal supergravity,''
  PTEP {\bf 2016}, no. 9, 093B03 (2016)
  %doi:10.1093/ptep/ptw119
  [arXiv:1606.06515 [hep-th]].  
  

\end{thebibliography}
\end{document}